\documentclass[review,authoryear,12pt]{elsarticle}

\usepackage{graphicx}
\usepackage[tight,footnotesize, nooneline]{subfigure}
\usepackage{adjustbox} % for squeezing tables
\usepackage[hidelinks]{hyperref} % pdf hyperlinks

\usepackage{amssymb}
\usepackage{amsthm}
\usepackage{amsmath}

\usepackage{lineno}
 
\usepackage{multirow}

\usepackage[flushleft]{threeparttable} %table footnotes

\usepackage{adjustbox} % for shrinking tables

\journal{arXiv}

\newcommand\copyrighttext{%
   \textcopyright\ 2020. This manuscript version is made available under the CC-BY-NC-ND 4.0 license \url{http://creativecommons.org/licenses/by-nc-nd/4.0/}}

\begin{document}

\begin{frontmatter}

%% Title


\title{Full feature selection for estimating KAP radiation dose in coronary angiographies and percutaneous coronary interventions}

%% Authors and addresses/affiliations

\author[address1]{Visa Suomi\corref{correspondingauthor}\fnref{myfootnote}}
\cortext[correspondingauthor]{Corresponding author}
%\ead{visa.suomi@utu.fi}
%\ead[url]{http://hifu.utu.fi/}
\author[address2]{Jukka J\"arvinen\fnref{myfootnote}}
\author[address3]{Tuomas Kiviniemi}
\author[address3]{Antti Ylitalo}
\author[address3]{Mikko Pietil\"a}

\address[address1]{Department of Radiology, Turku University Hospital, Kiinamyllynkatu 4-8, 20521 Turku, Finland}
\address[address2]{Department of Medical Physics, Turku University Hospital, Kiinamyllynkatu 4-8, 20521 Turku, Finland}
\address[address3]{Department of Cardiology, Turku University Hospital, Kiinamyllynkatu 4-8, 20521 Turku, Finland}

\fntext[myfootnote]{These authors contributed equally to this work.}

\begin{abstract}
In interventional cardiology (IC) the radiation dose variation is very significant, and its estimation has been difficult due to the complexity of the treatments. In order to tackle this problem, the aim of this study was to identify the most important demographic and clinical features to estimate Kerma-Area Product (KAP) radiation dose in coronary angiographies (CA) and percutaneous coronary interventions (PCI). The study was retrospective using clinical patient data from 838 CA and PCI procedures. A total of 59 features were extracted from the patient data and 9 different filter-based feature selection methods were used to select the most informative features in terms of the KAP radiation dose from the treatments. The selected features were then used in a support vector regression (SVR) model to evaluate their performance in estimating the radiation dose. The ten highest-ranking features were: 1) FN1AC (CA), 2) FN2BA (PCI), 3) weight, 4) post-stenosis 0\%, 5) multi-vessel disease, 6) number of procedures 3, 7) pre-stenosis 100\%, 8) American Heart Association (AHA) score C, 9) pre-stenosis 85\% and 10) gender. The performance of the SVR model increased (mean squared error $\approx$ 450) with the number of features approximately up to 30 features. The identification of the most informative features for CA and PCI KAP is an important step in determining suitable complexity models for clinical practice. The highest-ranking features can be used as individual predictors of IC procedure KAP or can be incorporated into combined complexity score or different estimation models in the future.
\end{abstract}

\begin{keyword}
machine learning \sep feature selection \sep radiation dose \sep coronary angiography \sep percutaneous coronary intervention
\end{keyword}

\end{frontmatter}

\copyrighttext

%% Do not remove the page break here.
\pagebreak

%\linenumbers

%% BEGIN MAIN TEXT

\section*{Introduction}

Angiographic imaging is a traditional and important part of radiation usage in medicine. In interventional cardiology (IC), angiographic imaging enables many life-saving and minimally invasive diagnostic and therapeutic procedures, and therefore, its usage has been growing fast and steadily over the recent years~\citep{kiviniemi1, kiviniemi2, kiviniemi3}. In addition to the obvious positive sides of the trend, there have also been concerns over the radiation dose levels in patients and staff members, which have potentially serious risks on both individual and population levels~\citep{icrp13}. In Finland roughly 29,000 coronary angiographies (CA) and 13,000 percutaneous coronary interventions (PCI) are conducted annually with the average Kerma-Area Product (KAP) dose levels being around 22 and 64~Gy$\cdot$cm$^{2}$~\citep{opas}, respectively. On a wider scale, approximately 1.8 million CA and 0.9 million PCI procedures were performed in Europe in 2015 according to the European Association of Percutaneous Cardiovascular Interventions (EAPCI) registry~\citep{barbato2017current}.

The level of radiation exposure is traditionally monitored using diagnostic reference levels (DRLs)~\citep{icrp135}, which are indicators for the typical dose levels for radiologic procedures. However, the application of DRLs in IC is somewhat challenging, because multiple factors affect the patient KAP dose significantly. Therefore, the utilisation of an estimate for procedure complexity to predict interventional use of radiation has been proposed by the International Commission on Radiological Protection (ICRP)~\citep{icrp135}. With accurate prediction, patients at risk of high radiation dose can be identified and their dose can be better optimised. However, the set of features that should be combined into such estimate remains an open question. In 2001 \citet{padovani2001patient} and in 2005 \citet{peterzol2005reference} published their pioneering works on the topic, but the only recent publication on features that predict high radiation dose is by \citet{crowhurst2019factors} who utilised a more conventional methodology. 

Elsewhere in IC, various difficulty or risk scores have been in use for a long time. For example, various scores are used to predict coronary disease risk or procedure outcome. These include American Heart Association/American College of Cardiology (AHA/ACC) lesion complexity~\citep{aha1, aha2, aha3} (aka AHA score), SYNTAX coronary disease complexity score~\citep{syntax} and European HeartScore for risk prediction~\citep{heartscore}, but none of these parameters focus specifically on the use of radiation. As contemporary practice is to begin the procedure with a diagnostic CA and, if needed, to directly continue with a therapeutic PCI, these scores can be considered to relate to both procedures.

In recent studies~\citep{siiskonen2018establishing, jukka} the radiation dose levels of various IC procedures have been reported. Based on the results, the variation in PCI doses between different countries and hospitals as well as within hospitals makes the interpretation and application of the DRLs relatively complicated. In addition, AHA/ACC score has only modest association with the induced radiation dose, which suggests that it is not a sufficient measure of PCI complexity in terms of KAP radiation dose~\citep{jukka}.

To address the need for a reliable indicator for contemporary IC procedure complexity, this study utilises multiple filter-based feature selection methods to identify the most informative patient demographic and clinical features for estimating the KAP radiation dose in CA and PCI procedures. Features related to performing cardiologist and use of the angiographic system were not included since these were not available in the dataset. The features chosen by the feature selection methods were then used in a supervised machine learning model to evaluate their performance in estimating the induced radiation dose. Beyond the benefits to scientific work on procedure complexity, the results of this study can be utilised to establish more accurate difficulty level-based DRLs and personalised dosimetry measures for patients at risk of high dose. 

\section*{Methods}

\subsection*{Data collection}

Anonymised patient data from CA and PCI procedures were collected retrospectively from the electronic dose records (DoseWatch, GEMS, Milwaukee, USA), PCI procedure registry (BCB, Turku, Finland) and electronic health records (Uranus, CGI, Montreal, Canada) at the Turku University Hospital, Finland. The data collection timeframe was two years from January 2016 to December 2017. The only patient inclusion criterion was the availability of information on both dose records and PCI procedure registry for each CA or PCI. The dose records collect IC procedure data automatically from the DICOM-file headers and image data whereas in PCI registry some of the data are manually entered by the practising cardiologist. Ethical permission for the usage of the patient data was received (T72/2016) from the Ethics Committee of Hospital District of Southwest Finland and the request for patient consent was waived due to the observational and anonymised nature of the study. A total of 838 procedures were included in the study of which 433 were CA and 405 were PCI. All the IC procedures were performed using the same angiography system (Artis Zee Ceiling, Siemens Healthineers, Erlangen, Germany) in the hospital.

\subsection*{Feature types}

The collected data included features relevant to patient demographics, disease status and the conducted angiographic procedure. Features that occurred in less than 1\% of the procedures were discarded to avoid misinterpretation of their importance. In total, 59 features were included in the analysis. Of the included features, 12 were considered PCI specific and 47 related to both CA and PCI. The features were not separated based on the procedure (i.e., CA or PCI) and calculations were thus performed with all of them. In order to avoid making a future complexity model overly complicated, the number of subgroups in the data was kept to as low as reasonably possible. To describe radiation dose to patient, KAP was chosen for its general application as an indicator of radiation dose, including its suitability for estimation of skin dose~\citep{jarvinen2018feasibility}.

Overviews of the collected features are presented in Tables~\ref{tab:datanum} and \ref{tab:datacat}. The last column in the tables indicates the ratio of missing values in the data. The missing manual inputs in binary features (yes/no) were interpreted as `no' and thus were not counted as missing. The radiation dose levels in patients ranged from 0.6 to 215.2~Gy$\cdot$cm$^{2}$ with the mean and standard deviation (SD) of 28.0 and 26.0~Gy$\cdot$cm$^{2}$, respectively.

\begin{table}[t!]
	\centering
	\begin{threeparttable}
	\caption{Numerical features from the interventions.}
		\begin{tabular}{lllllll}
		\hline
		\hline
		Feature 				& Unit					& Mean 		& SD 		& Min		& Max 	& \% missing 	\\
		\hline
		Age 					& (years) 			& 69.8 		& 11.7 	& 33.0	& 97.0 	& 0.0 				\\
		Weight 				& (kg) 				& 82.3		& 16.8 	& 44.0 	& 176.0 	& 0.0 				\\
		Height 				& (cm) 				& 171.3	 	& 9.3 		& 142.5	& 194.0 	& 27.1 			\\ 
		BSA\tnote{a}		& (m$^{2}$) 		& 1.9 			& 0.2 		& 1.4		& 2.8 		& 27.1 			\\
		BMI\tnote{b}		& (kg/m$^{2}$)	& 28.2 		& 4.9 		& 18.3	& 50.0 	& 27.1 			\\
		Stent dimension 	& (mm) 				& 3.3 			& 0.6 		& 2.3 		& 5.6 		& 74.9 			\\
		Ball dimension 		& (mm) 				& 3.2 			& 0.8 		& 1.0 		& 5.2 		& 84.1 			\\
		\hline
		\hline
		\end{tabular}
	\label{tab:datanum}
	\begin{tablenotes}\footnotesize
	\item[a] Body surface area
	\item[b] Body mass index
	\end{tablenotes}
	\end{threeparttable}
\end{table}

\begin{table}[htbp!]
	\centering
	\begin{adjustbox}{width=\columnwidth,center}
	\begin{threeparttable}
	\caption{Categorical features from the interventions.}
		\begin{tabular}{lll}
		\hline
		\hline
		Feature 							& Categories 													& \% missing 	\\
		\hline
		Gender 							& male/female 													& 0.0 			\\
		Procedure 						& FN1AC\tnote{a}/FN2BA\tnote{b}						& 0.0 			\\ 
		Indication 						& PCI in STEMI, Flap failure, NSTEMI, Diagnostic,		& 19.9 		\\
											& UAP, Heart failure, STEMI other, Stable AP or 		& 				\\  
											& Arrhythmia settlement 										& 				\\
		Coronary segment				& LADa, LADb, LADc, LCXa, LCXb, LCXc, LD1,			& 35.7 		\\
											& LM, LOM1, RCAa, RCAb, RCAc, RPD						& 				\\
		Diagnosis						& I35.0\tnote{c}, I20.81\tnote{d}, I21.01\tnote{e}, I21.11\tnote{f} or I21.41\tnote{g}																						& 0.0 			\\ 
		Multi-vessel disease 			& yes/no 														& 0.0 			\\ 
		Previous CABG\tnote{h} 		& yes/no 														& 0.0 			\\ 
		AHA/ACC score\tnote{i}		& A, B1, B2 or C 												& 50.2 		\\ 
		CTO\tnote{j}					& yes/no 														& 0.0 			\\ 
		Restenosis  						& yes/no 														& 0.0 			\\ 
		LM unprotected 				& yes/no 														& 0.0 			\\ 
		Pre-stenosis 					& 60\%, 85\% or 100\% 									& 51.7 		\\ 
		Post-stenosis 					& 0\%, 25\%, 60\%, 85\% or 100\% 					& 51.7 		\\ 
		Additional stenting 			& 1/over 1 														& 0.0 			\\ 
		N of procedures 				& 1, 2 or 3 														& 0.0 			\\ 
		\hline
		\hline
		\end{tabular}
		\label{tab:datacat}
		\begin{tablenotes}\footnotesize
		\item[a] Coronary Angiography (CA)
		\item[b] Percutaneous Coronary Intervention (PCI) with stent
		\item[c] Nonrheumatic aortic (valve) stenosis
		\item[d] Angina pectoris
		\item[e] ST elevation myocardial infarction (STEMI) involving left main coronary artery
		\item[f] STEMI involving right coronary artery
		\item[g] Myocardial infarction without ST elevation
		\item[h] Coronary artery bypass grafting
		\item[i] American Heart Association/American College of Cardiology
		\item[j] Chronic total occlusion
		\end{tablenotes}
	\end{threeparttable}
	\end{adjustbox}
\end{table}

The procedure-specific features were FN1AC denoting coronary angiography (CA) and FN2BA denoting PCI with stent. The diagnoses were nonrheumatic aortic (valve) stenosis (I35.0), angina pectoris (I20.81), ST elevation myocardial infarction (STEMI) involving left main coronary artery (I21.01), STEMI involving right coronary artery (I21.11) and myocardial infarction without ST elevation (I21.41). 

Within the indication-specific features, UAP denotes unstable angina pectoris. Segment refers to location of the disease and multi-vessel disease to multiple diseased segments. Previous CABG refers to previous coronary artery bypass grafting. The number (N) of procedures denotes number of billed procedures performed. Pre-stenosis refers to stenosis before PCI and post-stenosis to residual stenosis after PCI. CTO refers to chronic total occlusion of coronaries and LM unprotected to unprotected left main coronary in the procedure. AHA score refers to AHA/ACC lesion classification. Additional stenting 1 and over 1 refer to the usage of two or more stents, respectively. In addition, two synthetic parameters were created for body surface area (BSA) calculated with Du Bois formula~\citep{dubois1916formula} and body mass index (BMI) using the standard formula. Ball and stent dimensions refer to the diameters of the equipment used in PCI.

\subsection*{Feature selection methods}

A total of 9 different filter-based (i.e., independent of the regression model) feature selection methods were used in the analysis: F-value regression (FREG), mutual information regression (MIR), SURF (SURF), SURFstar (SURFS), MultiSURF (MSURF), MultiSURFstar (MSURFS), Pearson correlation coefficient (PEAR), Spearman's correlation coefficient (SPEA) and ReliefF (RELF). In addition, the aggregate rankings from all feature selection methods are referred to as TOPN. All the feature selection methods are publicly available from scikit-learn and scikit-rebate open access repositories~\citep{pedregosa2011scikit, urbanowicz2018benchmarking}. These feature selection methods were chosen because of their popularity in literature, computational efficiency and applicability for regression targets. Furthermore, the chosen methods have the advance of outputting individual feature weights which can be used for ordering features according to their importance for the specific target. More detailed descriptions about the feature selection methods used in this study can be found from the references~\citep{pedregosa2011scikit, urbanowicz2018benchmarking}.

\subsection*{Regression model and hyperparameter search}

\begin{figure}[b!]
\centering
\includegraphics[width=1\columnwidth]{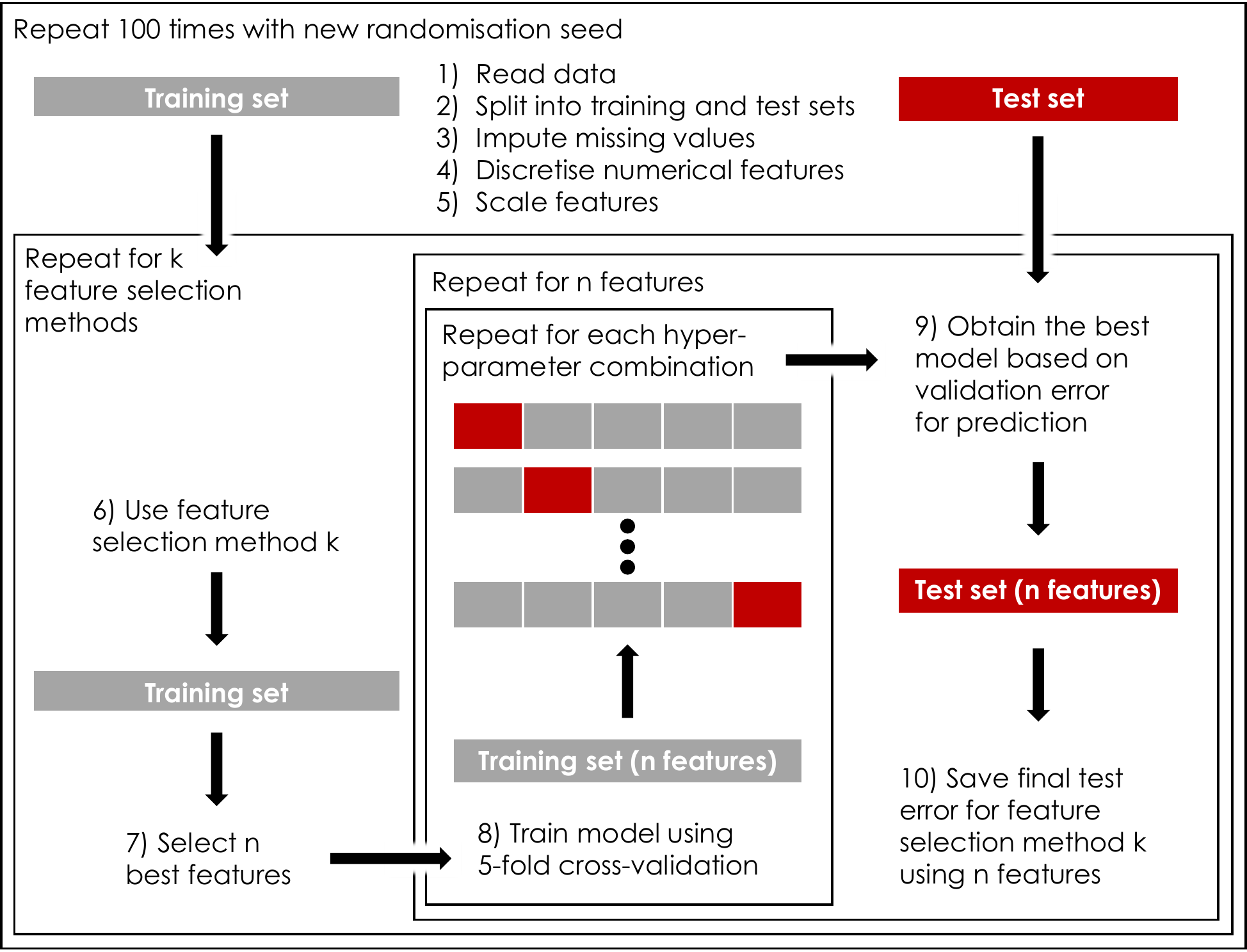}
\caption{Overview of the data processing pipeline: 1) The data was read into a dataframe; 2) split into training and test sets; 3) imputed based on the training set; 4) numerical features were discretised; 5) all features were log-scaled; 6) feature selection method ($k$ = 1-9) was used on the training set; 7) the highest-ranking features ($n$ = 5, 10,..., 40) were obtained; 8) the $n$ features from the method $k$ were used to train a support vector regression (SVR) model using hyperparameter grid search with inner 5-fold cross-validation; 9) the SVR model was refit on the whole training set using the combination of hyperparameters based on the lowest cross-validation error (mean squared error, MSE); 10) the fitted SVR model was used to predict the radiation dose in the test set with the same $n$ features and the test error (MSE) was saved. Steps 1-10) were repeated 100 times with a new randomisation seed.}
\label{fig:algorithm_pipeline}
\end{figure}

An overview of the data processing pipeline is presented in Figure~\ref{fig:algorithm_pipeline}. The methodology is similar to the previous study~\citep{suomi2019comprehensive} but with the difference that the target is continuous in this case rather than a class. The dataset of 838 patients was first read into a dataframe and was randomly split with 80\% for training and 20\% for testing. The features with missing values were imputed using their mean for numerical and a constant value (0) for categorical features. The mean imputation values were based only on the training data in order to avoid any information leakage from the test data. 

The numerical features were discretised (ordinal) into ten equal width bins and the categorical features were one-hot encoded. All features were then logarithmically scaled in order to optimise the regression model performance. Once the data were pre-processed, the feature selection methods ($k$ = 1-9) were used on the training set to rank the $n$ best features ($n$ = 5, 10,..., 40). These $n$ features from the training set were then used as inputs for the regression model.

Support vector regression (SVR) model was selected as the regression method for the problem since it does not have any built-in feature selection methods (i.e., embedded or wrapper). SVR is a supervised machine learning model whose basic idea is to predict continuous output values (i.e., real numbers) by minimising the pre-defined loss (error) function. In this study mean squared error (MSE) was selected as the loss function. A margin of tolerance $\varepsilon$ is also determined and no penalty is associated in the loss function with points predicted within a distance $\varepsilon$ from the actual value. In addition, since SVR does a linear prediction on the data, a kernel function was used to transform the data into a higher dimensional feature space to make it possible to perform the linear separation. In this study, nonlinear radial basis function (RBF) kernel was used because it was found to perform better with the given dataset compared to linear and polynomial kernel models. The SVR model was implemented using scikit-learn (v0.20.3) in Python (3.5.3)~\citep{pedregosa2011scikit}.

Before starting the SVR model fitting process with the selected $n$ features from the training data, a set of hyperparameters were determined for grid search (see Table~\ref{tab:hyperparameters}). The aim of the grid search was to find the optimal combination of hyperparameters for each set of features by iterating over all the possible combinations of the given grid. The performance of each hyperparameter combination was evaluated using 5-fold cross-validation on the training data and calculating a validation error (MSE) for each fold. Using 5-folds was found sufficient so that both sets represented the underlining distribution of the data. The mean error over all 5 folds was then selected as the performance metric of the given hyperparameter combination and the process was repeated for the next set of hyperparameters. Once all possible hyperparameter combinations had been evaluated, the lowest mean error was selected as the optimal set of hyperparameters for the given $n$ features. The SVR model was then refitted on the whole training set using these hyperparameters. Finally, the fitted model was used to make predictions on the test set with the same $n$ features and the test error (MSE) was calculated based on the test predictions as the final performance metric.

The whole process above was repeated 100 times using a new randomisation seed in every iteration, which was found to ensure the stability and repeatability of the results. All the results are therefore average values over 100 iterations. The total computation time was approximately 18 hours using 24 CPU units (Intel Xeon E5-2670 @ 2.60~GHz) on a computing cluster.

\begin{table}[t!]
	\centering
	\begin{threeparttable}
  	\caption{Hyperparameters for grid search in support vector regression (SVR) model fitting.}
    	\begin{tabular}{ll}
    	\hline
    	\hline
   	Parameter 				& Values  							\\
    	\hline
    	Kernel 					& RBF\tnote{a} 					\\
    	$\varepsilon$			& 0.1 									\\
    	C     						& 1, 1e1, 1e2, 1e3, 1e4, 1e5 	\\
    	Gamma 					& 1e-1, 1, 1e1, 1e2, 1e3 		\\
    	\hline
    	\hline
    \end{tabular}
    \label{tab:hyperparameters}
    \begin{tablenotes}\footnotesize
		\item[a] Radial Basis Function
		\end{tablenotes}
	\end{threeparttable}
\end{table}

\section*{Results}

\subsection*{Feature importances and correlation}

\begin{figure}[b!]
\centering
\includegraphics[width=1\columnwidth]{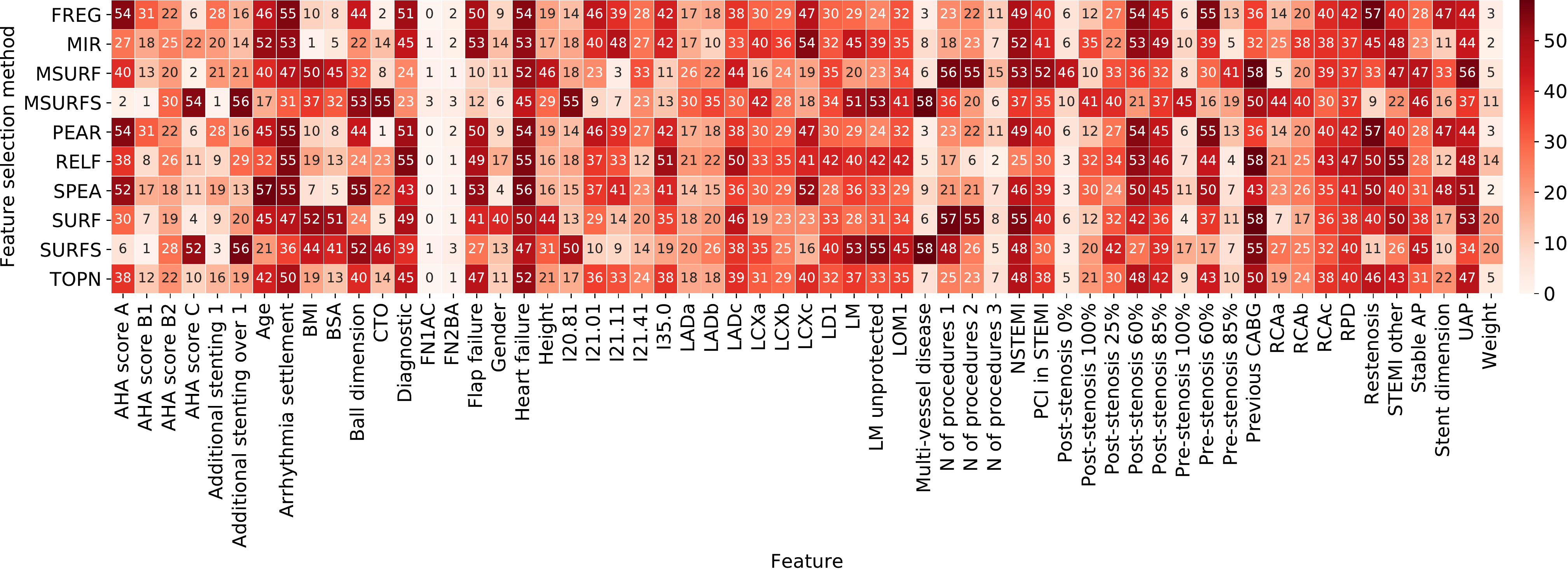}
\caption{Heatmap showing the median feature rankings (range 0-58 from the best to worst) from 9 different filter-based feature selection methods (100 rankings per feature) in estimating the procedure radiation dose. In addition, the median ranking from all methods (9 methods $\times$ 100 repetitions = 900 rankings per feature) for each feature is shown at the bottom (TOPN).}
\label{fig:heatmap_rankings_median}
\end{figure}

A heatmap of the median feature rankings (range 0-58 from the best to worst) in estimating the procedure radiation dose is shown in Figure~\ref{fig:heatmap_rankings_median}. The rankings are shown for all 59 features from the 9 different feature selection methods. Each feature had 100 rankings per method and their corresponding median value is presented in the map. In addition, the aggregate median rankings from all methods (9 methods $\times$ 100 repetitions = 900 rankings per feature) are shown at the bottom (TOPN).

\begin{figure}[b!]
\centering
\includegraphics[width=1\columnwidth]{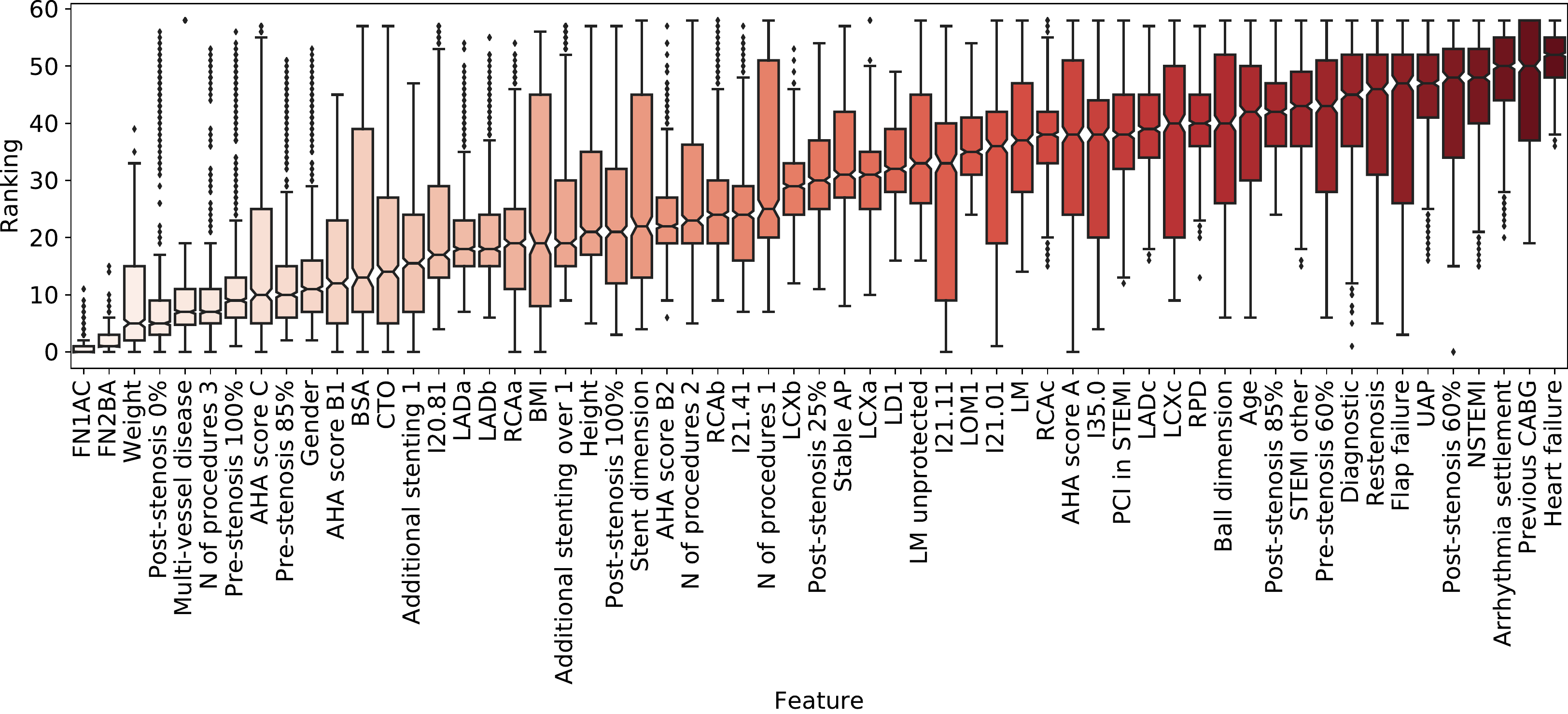}
\caption{Boxplot showing the feature rankings (range 0-58 from best to worst) from 9 different filter-based feature selection methods (9 methods $\times$ 100 repetitions = 900 rankings per feature) in estimating the procedure radiation dose. The features are ordered by their median value based on the rankings. Boxes show the interquartile ranges (IQR) with median values (notch) and whiskers show 1.5 IQR from the lower and upper quartiles. Outliers are plotted as individual points beyond the ends of the whiskers.}
\label{fig:boxplot_feature_rankings}
\end{figure}

\begin{figure}[b!]
    \centering
    \subfigure[]
    {
        \includegraphics[height=6cm]{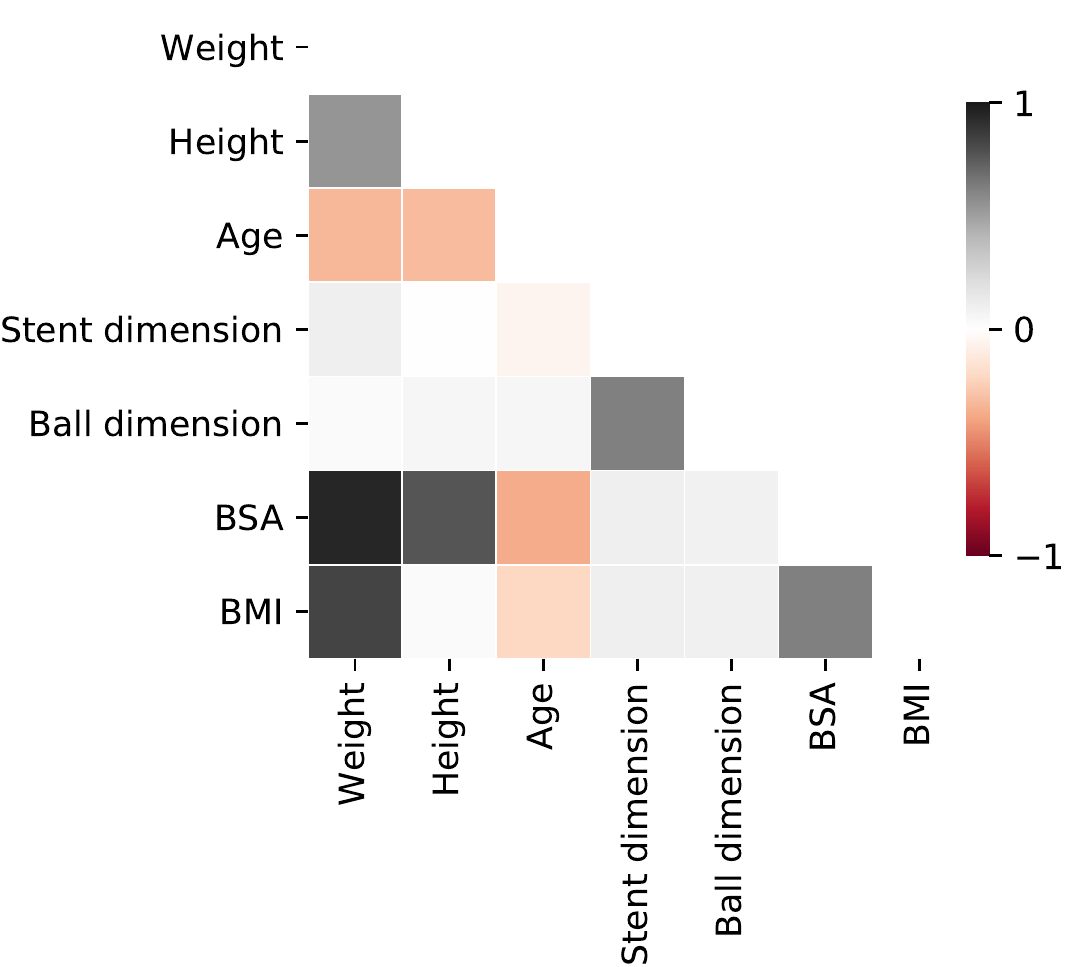}
    }
    \subfigure[]
    {
        \includegraphics[height=6cm]{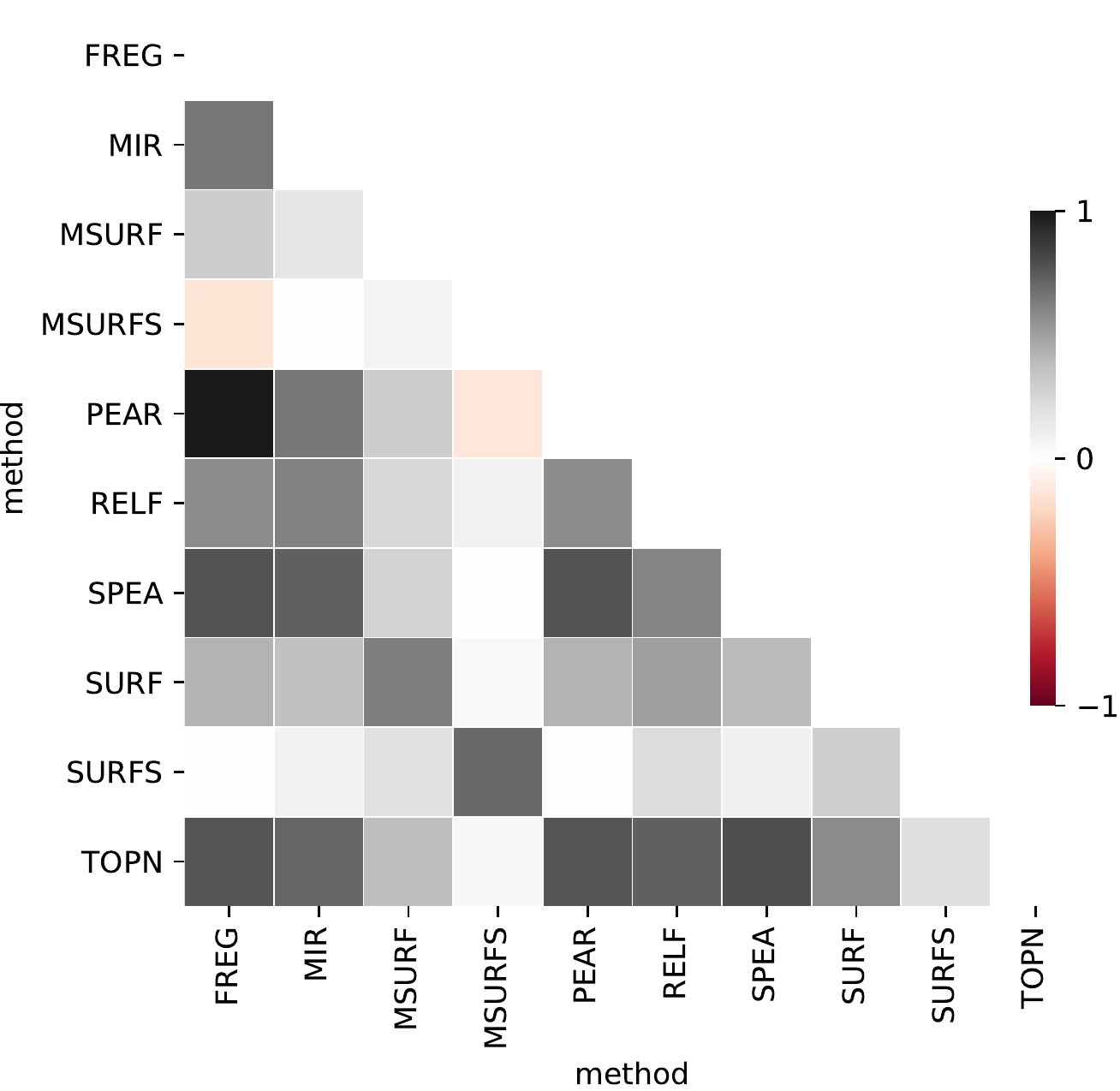}
    }
\caption{Diagonal correlation matrices showing (a) the pairwise Spearman's correlation coefficients between numerical features and (b) Kendall's tau rank correlation coefficient showing the similarity between different feature selection methods in ranking the features.}
\label{fig:feature_corr}
\end{figure}

The statistical distributions of the aggregate rankings are visualised in the boxplot in Figure~\ref{fig:boxplot_feature_rankings} which shows the features ordered by their median rankings (i.e., TOPN in Figure~\ref{fig:heatmap_rankings_median}). The boxes display the interquartile ranges (IQR) and the median values are marked with a notch inside each box. The whiskers show 1.5 IQR from the lower and upper quartiles and outliers are plotted as individual points beyond the ends of the whiskers. In addition, the pairwise Spearman's correlation coefficients between numerical features and Kendall's tau rank correlation coefficients between different feature selection methods are shown in Figures~\ref{fig:feature_corr}(a) and (b), respectively.

The ten highest-ranking features organised by their aggregate median values were: 1) FN1AC, 2) FN2BA, 3) weight, 4) post-stenosis 0\%, 5) multi-vessel disease, 6) N of procedures 3, 7) pre-stenosis 100\%, 8) AHA score C, 9) pre-stenosis 85\% and 10) gender. Some of the features have relatively large IQRs and whiskers, which indicates a certain degree of disagreement between the methods. There are also visible outliers outside the whiskers in some of the features.

The rankings shown in Figures~\ref{fig:heatmap_rankings_median} and \ref{fig:boxplot_feature_rankings} illustrate the significance of diagnostic vs. therapeutic as FN1CA was ranked the most important and FN2BA the second. This result is due to the significant additional dose from the intervention compared to the diagnostic CA~\citep{stukdrl}. Patient weight is a traditional radiation dose predictor and its high rank was also expected. 

Post-stenosis 0\% is associated with higher KAP (mean 36.7~Gy$\cdot$cm$^{2}$ compared to mean 28.0~Gy$\cdot$cm$^{2}$ for the whole data) and was ranked fourth. This association can be interpreted to be due to required finesse and subsequent imaging to achieve said result. Multi-vessel disease and the high number of performed procedures (3) ranked fifth and sixth, respectively. They are associated with higher radiation dose as they directly increase the extent and urgency of the procedure. High number of procedures is strongly associated with PCI, which likely influences and artificially increases its rank. Pre-stenosis 100\%, AHA score C and pre-stenosis 85\% were ranked seventh, eighth and ninth, respectively. They all imply difficult procedure with high pre-stenosis and tortuous or degenerated vein grafts. 

Patient gender was ranked tenth with men induced to 32.5~Gy$\cdot$cm$^{2}$ dose on average compared to 18.8~Gy$\cdot$cm$^{2}$ in women. This difference is partly attributable to associations with patient weight, additional stenting over 1, multivessel disease, number of procedures 3 and PCI, which were all more common with men. These results point to men in the study seeking treatment at a later stage of disease.

AHA score B1 ranked 11th but B2 and A ranked significantly lower. The AHA score takes into account the lesion location, bifurcation, size and shape and coronary tortuosity and angulation. These factors describe PCI complexity in terms of outcome and the result indicates that its applicability to procedure difficulty in terms of use of radiation is not straightforward.

BSA ranked 12th most likely affected by its high correlation to weight. CTO ranked 13th, which is lower than expected. The result is likely affected by the low number of CTOs in the data (32 samples) and huge dose variation among them (mean dose 69.4~Gy$\cdot$cm$^{2}$ and standard deviation 56.3~Gy$\cdot$cm$^{2}$. Additional stenting 1 and over 1 ranked 14th and 20th, respectively. Additional stenting implies difficulty in performing the procedure and need for additional imaging. 

The relatively high number of missing values in AHA scores (50.2\% missing), pre-stenosis and post-stenosis (51.7\% missing) values indicates that these results might not be as accurate as for the rest of the features. Post-stenosis 0\% (fourth), pre-stenosis 100\% (seventh), AHA score C (eighth) and pre-stenosis 85\% (ninth) all ranked in the top ten despite of the missing data which could indicate that there is still a strong correlation between these features and the KAP radiation dose despite of some of the values missing. 

\subsection*{Prediction performance}

Once the most important features were determined using all the feature selection methods, their performance in estimating the radiation dose was evaluated with the SVR model. For this purpose, the number of input features for the model was varied between 5 and 40 using the highest-ranking features from each method. The maximum number of features included in the prediction was stopped at 40 because the prediction performance did not improve considerably after this point. 

\begin{figure}[b!]
    \centering
    \subfigure[]
    {
        \includegraphics[height=4.6cm]{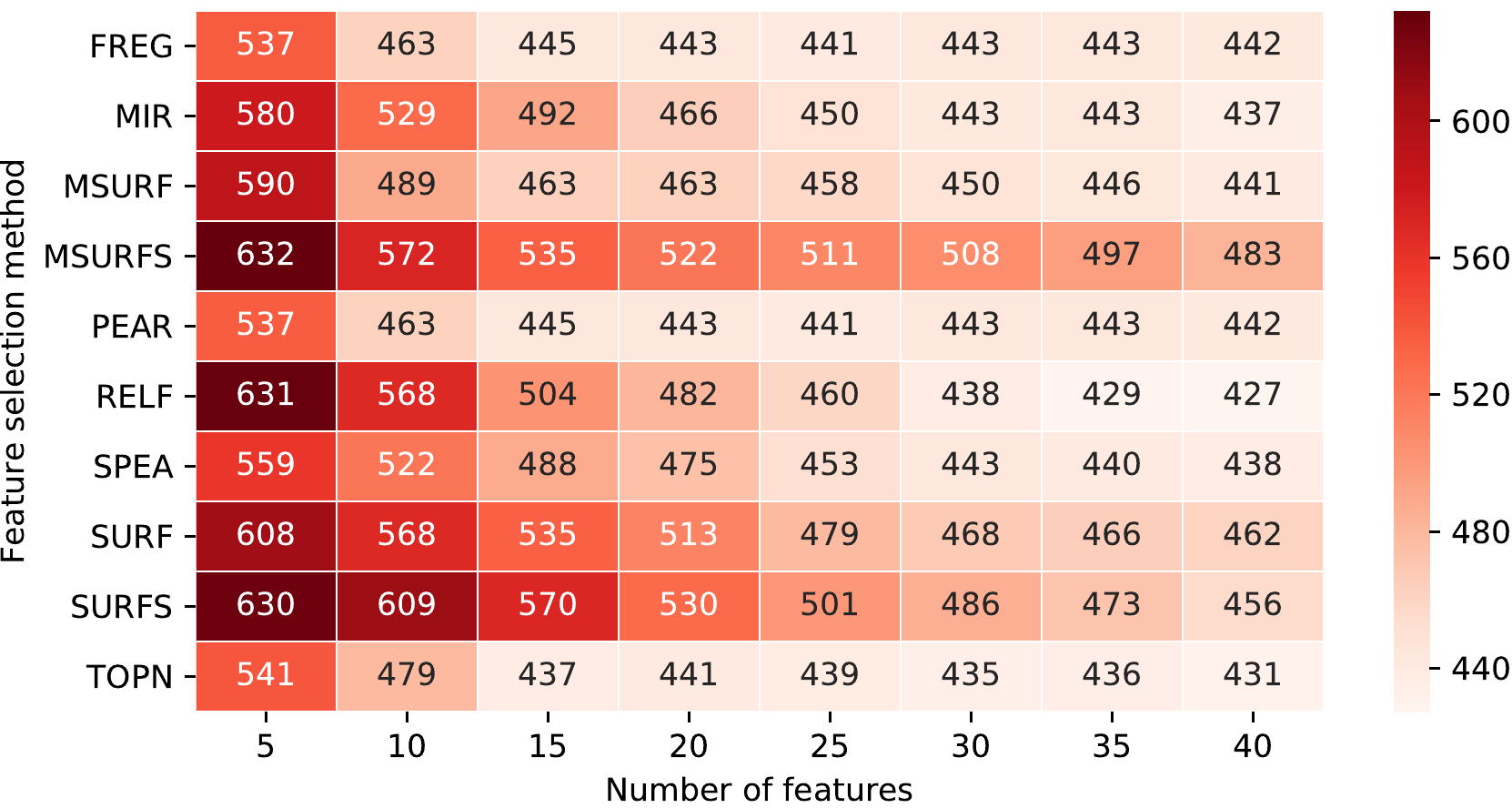}
    }
    \subfigure[]
    {
        \includegraphics[height=4.6cm]{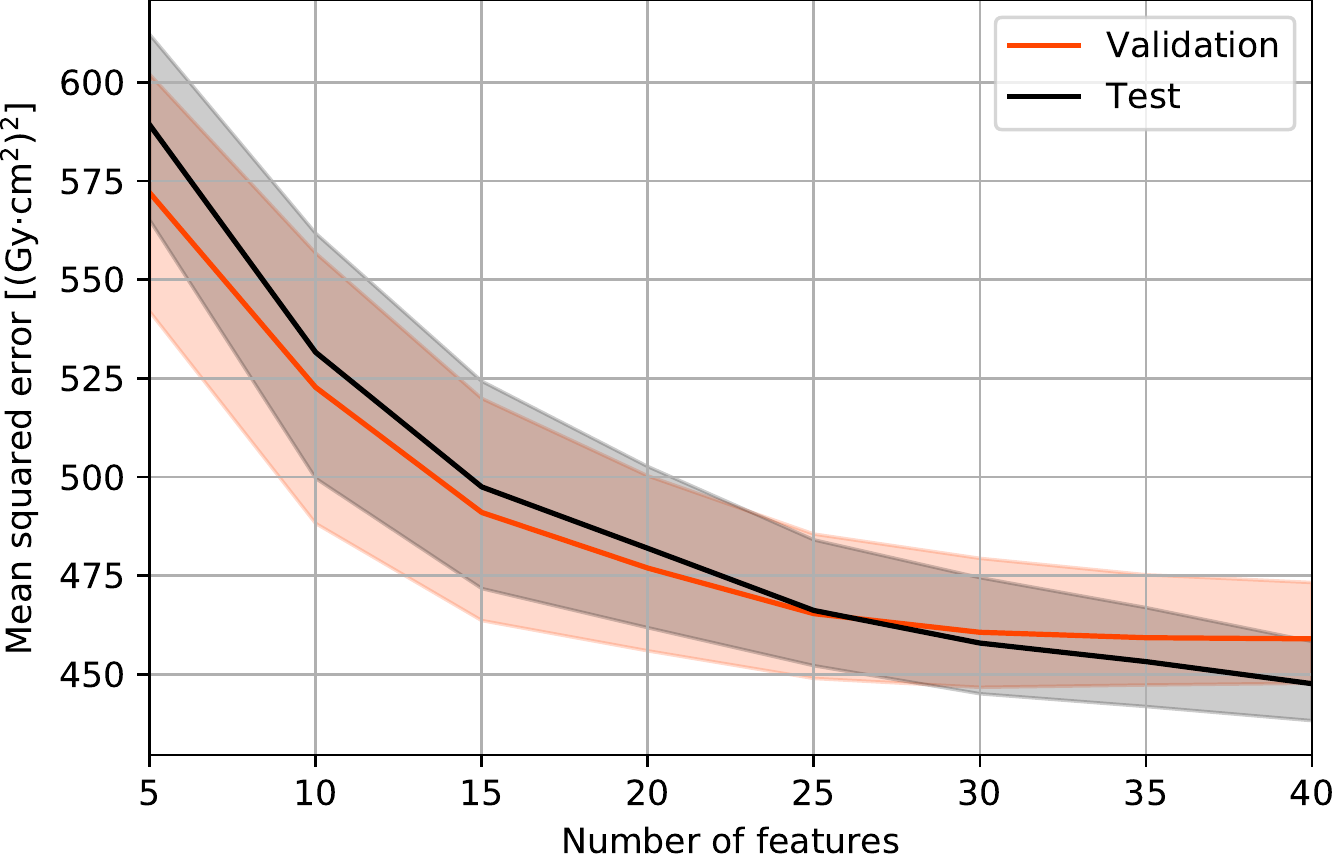}
    }
    \caption{(a) Heatmap showing the mean test errors from support vector regression (SVR) model in estimating the procedure radiation dose. The values show the mean test error from 100 repetitions using the stated number of highest-ranking features from each feature selection method. In addition, the mean test errors using the highest-ranking features from aggregate votes are shown at the bottom (TOPN). (b) Lineplot showing the mean and dispersion of validation and test errors with the number of highest-ranking features from all feature selection methods. The faded areas show the 95\% confidence intervals.}
    \label{fig:cross-validation_errors}
\end{figure}

Figure~\ref{fig:cross-validation_errors}(a) shows a heatmap of the average test errors over 100 iterations using the specified number of highest-ranking features from each feature selection method. In addition, the mean test errors using the highest-ranking features from aggregate votes (see Figure~\ref{fig:boxplot_feature_rankings}) are shown at the bottom (TOPN). Figure~\ref{fig:cross-validation_errors}(b) shows the mean and dispersion of the validation and test errors as a function of highest-ranking features from all feature selection methods. The faded areas show the 95\% confidence intervals.

Both the mean validation and test errors were reduced when the number of input features for the SVR model was increased. However, there are differences in the performance of different feature selection methods. FREG, MIR, MSURF, PEAR, RELF and SPEA appear to have approximately similar test errors (MSE range 427-442) with the maximum number of features (40). SURF and SURFS perform slightly worse (MSE range 456-462) compared to the former methods while MSURFS has the worst performance (MSE 483). This indicates that the features selected by the methods with high test errors are not as informative in estimating the treatment radiation dose as the ones chosen by methods with lower test errors. However, this does not mean that the worse performing methods are inferior \textit{per se}, but rather tells about their suitability for the given problem.

The second-best dose estimation performance (after RELF) was achieved using the top features from the aggregate rankings (MSE 431). This indicates that these features are indeed informative in estimating the treatment radiation dose and should be included as input features for the prediction model. However, the optimal number of input features still has to be determined. In Figure~\ref{fig:cross-validation_errors}(b) the mean validation and test errors are rapidly decreased when the number of features is increased. The decrease flattens at around 30 features and only slight performance increases are seen when the feature set is grown beyond this number. Therefore, using the first 30 features from the aggregate rankings (see Figure~\ref{fig:boxplot_feature_rankings}) are enough to yield good balance between the number of features and dose estimation performance.

The accuracy of the SVR model can be estimated by examining the square root of the MSE in the results in Figure~\ref{fig:cross-validation_errors}. Therefore, the best accuracy with the given features was around MSE 427~(Gy$\cdot$cm$^{2}$)$^{2}$ (i.e. using RELF feature selection method and 40 features) whose square root is approximately 20.7~Gy$\cdot$cm$^{2}$. This gives an estimate of the mean error when the model is making predictions about the radiation dose. The error might seem large compared to the mean radiation dose of 28.0~Gy$\cdot$cm$^{2}$ in the data, but it is also affected by the large spread in KAP values from 0.6 to 215.0~Gy$\cdot$cm$^{2}$ all of which contribute to the mean error. Furthermore, the main aim of the study was not to obtain the best possible dose estimation accuracy from the model but to identify the features which are the most relevant in estimating the radiation dose. Therefore, the features identified in this study can be incorporated in future studies for developing a more accurate prediction model.

\section*{Discussion}

The results of this study show that the accuracy in estimating the radiation dose during CA and PCI procedures increases with the number of features. SVR showed increase in performance (i.e., lower MSE) all the way up to approximately 30 features, which indicates that these features contain predictive information about the induced radiation dose and could therefore be used as reliable indicators for the radiation dose as well as IC procedure complexity. The features beyond this point most likely do not play an important role in contributing to the overall radiation dose to the patient. However, it should be noted that the features were ranked according to their overall correlation with the radiation dose, and therefore, these features act either as negative or positive indicators for the radiation dose.

In terms of the prediction model accuracy, the ten highest-ranking features cannot be interpreted as the absolute truth for the optimal prediction performance, but rather as a guideline of which features should be considered in order to build accurate prediction models to estimate KAP radiation dose. The ranking of these features was based on the dataset used in the study, and therefore, comparison to multi-centre studies should be made in order to decide the most informative features for the given problem.

It was also evident from the results that many patient demographic and clinical features predict usage of radiation in IC and division to CA and PCI, AHA score or patient weight alone are not enough to estimate it accurately. In addition to its subjectivity and inaccuracy of the scoring~\citep{KleinAngiographicCO}, the relevance of the features in AHA scores for the use of radiation is relatively complex. The issue with patient weight in IC is quite well established by the ICRP~\citep{icrp135}. 

Besides these three features, multi-vessel disease, high pre-stenosis, gender, CTO~\citep{sakano2017survey, siiskonen2018establishing} or additional stenting should be considered due to their clear relevance to the use of radiation. Despite its strong association with PCI, high number of procedures is also possibly an important feature.

Compared to the recent results on factors contributing to high radiation dose in CA published by \citet{crowhurst2019factors}, the assessed features were different but similarities in the results can nonetheless be found. The most important statistically significant factors reported by Crowhurst et al. were patient weight, PCI and gender. In addition, statistically significant effect was also found for coronary artery bypass graft angiography, use of optical coherence tomography (OCT), intravascular ultrasound (IVUS) or fractional flow reserve (FFR), right heart catheterisation with or without venography, operator experience, and the type of the angiography system. 

The results also agree reasonably well with the earlier estimations of features predicting radiation dose and difficulty levels by \citet{padovani2001patient} and \citet{peterzol2005reference}. Padovani et al. identified the number of lesions, bifurcation stents, ostial stents, occlusions of older than three months and severe coronary tortuosity. Peterzol et al. identified complex lesions, double wire techniques, double balloon techniques, bifurcation stents and severe tortuosity in their model. Compared to these models, the results presented in this paper allow for significant updates better corresponding to contemporary practice.

The highest-ranking features provide guidance into how to set up estimates of procedure complexity from the perspective of use of radiation. However, to account for hospital and cardiologist specific variations a multi-centre cohort study is required. In addition, the applied methodology can be utilised for other procedures such as transcatheter aortic valve implantations and the optimisation of the use of radiation as to which practices increase radiation dose the most and what patient demographic and clinical features affect these practices. Therefore, the results of this study are considered as an essential part of a roadmap to more accurate difficulty level based DRLs and personalised dosimetry.

Limitations of the study include the single-centre data, the amount of missing data in some of the features, and the possibility of human errors especially in the PCI registry records due to manual input. In addition, features related to the cardiologist and the use of the angiographic system can be interpreted as confounders. These include factors such as the time of the procedure (e.g. morning, night) as well as the fatigue and experience level of the cardiologist all of which affect the procedure length, and therefore, also the radiation dose of the patient. These limitations could be corrected with a multi-centre study with these features included together with enough data from each hospital. For the purpose of meeting the aims of the present study, the amount of data can be considered to be sufficient for all the high-ranking features.

\section*{Conclusions}

Filter-based feature selection methods together with a supervised machine learning model were employed to determine the most informative features in estimating the radiation dose in CA and PCI procedures. The features most informative of high dose were identified to be PCI, high weight, high AHA score, multi-vessel disease, high pre-stenosis, male gender, CTO and additional stenting. The features were evaluated in terms of their ability to estimate the induced radiation dose, and after multi-centre validation and careful clinical assessment can be used to predict IC coronary procedure complexity from the perspective of use of radiation. Furthermore, the identification of the most informative features will advance the development and adaptation of similar machine learning models into clinical practice.

\section*{Acknowledgements}
\addcontentsline{toc}{section}{\numberline{}Acknowledgements}

V.~S. acknowledges the support of the State Research Funding (ERVA), Hospital District of Southwest Finland, number K3007 and CSC - IT Center for Science, Finland, for computational resources. J.~J. acknowledges the important work by Saija Hurme from Department of Biostatistics, University of Turku, Finland, in carrying out the univariate testing. 

%% Do not remove the page break here.
\pagebreak

%% References with bibTeX database, use this to create a .bbl file
\bibliographystyle{UMB-elsarticle-harvbib}
\bibliography{arXiv_computers_biology_medicine}

\begin{thebibliography}{25}
\expandafter\ifx\csname natexlab\endcsname\relax\def\natexlab#1{#1}\fi
\expandafter\ifx\csname url\endcsname\relax
  \def\url#1{\texttt{#1}}\fi
\expandafter\ifx\csname urlprefix\endcsname\relax\def\urlprefix{URL }\fi

\bibitem[{Barbato et~al.(2017)Barbato, Dudek, Baumbach, Windecker, and
  Haude}]{barbato2017current}
Barbato E, Dudek D, Baumbach A, Windecker S, Haude M. Current trends in
  coronary interventions: an overview from the eapci registries.
  EuroIntervention: journal of EuroPCR in collaboration with the Working Group
  on Interventional Cardiology of the European Society of Cardiology,
  2017;13:Z8--Z10.

\bibitem[{Blackledge and Squire(2009)}]{kiviniemi2}
Blackledge HM, Squire IB. Improving long-term outcomes following coronary
  artery bypass graft or percutaneous coronary revascularisation: results from
  a large, population-based cohort with first intervention 1995--2004. Heart,
  2009;95:304--311.

\bibitem[{Conroy et~al.(2003)Conroy, Py\"or\"al\"a, Fitzgerald, Sans, Menotti,
  De~Backer, De~Bacquer, Ducimetiere, Jousilahti, Keil, Nj$\o$lstad, Oganov,
  Thomsen, Tunstall-Pedoe, Tverdal, Wedel, Whincup, Wilhelmsen, Graham, and
  {SCORE project group}}]{heartscore}
Conroy RM, Py\"or\"al\"a K, Fitzgerald AP, Sans S, Menotti A, De~Backer G,
  De~Bacquer D, Ducimetiere P, Jousilahti P, Keil U, Nj$\o$lstad I, Oganov RG,
  Thomsen T, Tunstall-Pedoe H, Tverdal A, Wedel H, Whincup P, Wilhelmsen L,
  Graham IM, {SCORE project group}. Estimation of ten-year risk of fatal
  cardiovascular disease in europe: the score project. European Heart journal,
  2003;24:987--1003.

\bibitem[{Crowhurst et~al.(2019)Crowhurst, Whitby, Savage, Murdoch, Robinson,
  Shaw, Gaikwad, Saireddy, Hay, and Walters}]{crowhurst2019factors}
Crowhurst JA, Whitby M, Savage M, Murdoch D, Robinson B, Shaw E, Gaikwad N,
  Saireddy R, Hay K, Walters DL. Factors contributing to radiation dose for
  patients and operators during diagnostic cardiac angiography. Journal of
  medical radiation sciences, 2019;66:20--29.

\bibitem[{DuBois(1916)}]{dubois1916formula}
DuBois D. A formula to estimate the approximate surface area if height and body
  mass be known. Arch Intern Med, 1916;17:863--871.

\bibitem[{Ellis et~al.(1988)Ellis, Roubin, King, Douglas, Weintraub, Thomas,
  and Cox}]{aha2}
Ellis SG, Roubin GS, King SB, Douglas JS, Weintraub WS, Thomas RG, Cox WR.
  Angiographic and clinical predictors of acute closure after native vessel
  coronary angioplasty. Circulation, 1988;88:372--379.

\bibitem[{Ellis et~al.(1990)Ellis, Roubin, King, Douglas, Weintraub, Thomas,
  and Cox}]{aha3}
Ellis SG, Roubin GS, King SB, Douglas JS, Weintraub WS, Thomas RG, Cox WR.
  Coronary morphologic and clinical determinants of procedural outcome with
  angioplasty for multivessel coronary disease. implications for patient
  selection. multivessel angioplasty prognosis study group. Circulation,
  1990;82:1193--1202.

\bibitem[{Fokkema et~al.(2013)Fokkema, James, Albertsson, Akerblom, Calais,
  Eriksson, Jensen, Nilsson, de~Smet, Sj{\"o}gren, et~al.}]{kiviniemi3}
Fokkema ML, James SK, Albertsson P, Akerblom A, Calais F, Eriksson P, Jensen J,
  Nilsson T, de~Smet BJ, Sj{\"o}gren I, et~al. Population trends in
  percutaneous coronary intervention: 20-year results from the scaar (swedish
  coronary angiography and angioplasty registry). Journal of the American
  College of Cardiology, 2013;61:1222--1230.

\bibitem[{ICRP(2013)}]{icrp13}
ICRP. Radiological protection in cardiology. {ICRP} publication 120. Annals of
  ICRP, 2013;42.

\bibitem[{ICRP(2017)}]{icrp135}
ICRP. Diagnostic reference levels in medical imaging. {ICRP} publication 135.
  Annals of ICRP, 2017;46.

\bibitem[{J\"arvinen et~al.(2018)J\"arvinen, Eskola, Hallinen, J\"arvinen,
  Kivel\"a, M\"akel\"a, Parviainen, Pirinen, Rissanen, Sierpowska, Siiskonen,
  and Vinni-Lappalainen}]{opas}
J\"arvinen H, Eskola M, Hallinen E, J\"arvinen J, Kivel\"a A, M\"akel\"a T,
  Parviainen T, Pirinen M, Rissanen T, Sierpowska J, Siiskonen T,
  Vinni-Lappalainen K. STUK Opastaa 2018: S\"ateilyn k\"ayt\"on turvallisuus
  kardiologiassa. STUK, 2018.

\bibitem[{Jarvinen et~al.(2018)Jarvinen, Farah, Siiskonen, Ciraj-Bjelac, Dabin,
  Carinou, Domienik-Andrzejewska, Kluszczynski, Kne{\v{z}}evi{\'c}, Kopec,
  et~al.}]{jarvinen2018feasibility}
Jarvinen H, Farah J, Siiskonen T, Ciraj-Bjelac O, Dabin J, Carinou E,
  Domienik-Andrzejewska J, Kluszczynski D, Kne{\v{z}}evi{\'c} {\v{Z}}, Kopec R,
  et~al. Feasibility of setting up generic alert levels for maximum skin dose
  in fluoroscopically guided procedures. Physica Medica, 2018;46:67--74.

\bibitem[{J\"arvinen et~al.(2019)J\"arvinen, Sierpowska, Siiskonen, J\"arvinen,
  Kiviniemi, Rissanen, Matikka, Niskanen, Hurme, Larjava, M\"akel\"a,
  Strengell, Eskola, Parviainen, Hallinen, Pirinen, Kivel\"a, and
  Ter\"as}]{jukka}
J\"arvinen J, Sierpowska J, Siiskonen T, J\"arvinen H, Kiviniemi T, Rissanen
  TT, Matikka H, Niskanen E, Hurme S, Larjava HRS, M\"akel\"a TJ, Strengell S,
  Eskola M, Parviainen T, Hallinen E, Pirinen M, Kivel\"a A, Ter\"as M.
  Contemporary radiation doses in interventional cardiology: A nationwide study
  of patient doses in finland. Radiation protection dosimetry, 2019.

\bibitem[{Kappetein et~al.(2006)Kappetein, Dawkins, Mohr, Morice, Mack,
  Russell, Pomar, and Serruys}]{syntax}
Kappetein AP, Dawkins KD, Mohr FW, Morice MC, Mack MJ, Russell ME, Pomar J,
  Serruys PWJC. Current percutaneous coronary intervention and coronary artery
  bypass grafting practices for three-vessel and left main coronary artery
  disease. insights from the syntax run-in phase. European journal of
  cardio-thoracic surgery, 2006;29:486--491.

\bibitem[{Kiviniemi et~al.(2016)Kiviniemi, Pietil{\"a}, Gunn, Aittokallio,
  M{\"a}h{\"o}nen, Salomaa, and Niiranen}]{kiviniemi1}
Kiviniemi TO, Pietil{\"a} A, Gunn JM, Aittokallio JM, M{\"a}h{\"o}nen M,
  Salomaa VV, Niiranen TJ. Trends in rates, patient selection and prognosis of
  coronary revascularisations in finland between 1994 and 2013: the {CVDR}.
  EuroIntervention, 2016;12:1117--1125.

\bibitem[{Lloyd and Ronald(2008)}]{KleinAngiographicCO}
Lloyd WK, Ronald JK. Angiographic characterization of lesion morphology are the
  aha / acc and scai lesion classifications still useful? Cardiac Interventions
  Today, 2008.

\bibitem[{Padovani et~al.(2001)Padovani, Bernardi, Malisan, Vano, Morocutti,
  and Fioretti}]{padovani2001patient}
Padovani R, Bernardi G, Malisan MR, Vano E, Morocutti G, Fioretti PM. Patient
  dose related to the complexity of interventional cardiology procedures.
  Radiation protection dosimetry, 2001;94:189--192.

\bibitem[{Pedregosa et~al.(2011)Pedregosa, Varoquaux, Gramfort, Michel,
  Thirion, Grisel, Blondel, Prettenhofer, Weiss, Dubourg,
  et~al.}]{pedregosa2011scikit}
Pedregosa F, Varoquaux G, Gramfort A, Michel V, Thirion B, Grisel O, Blondel M,
  Prettenhofer P, Weiss R, Dubourg V, et~al. Scikit-learn: Machine learning in
  python. Journal of machine learning research, 2011;12:2825--2830.

\bibitem[{Peterzol et~al.(2005)Peterzol, Quai, Padovani, Bernardi, Kotre, and
  Dowling}]{peterzol2005reference}
Peterzol A, Quai E, Padovani R, Bernardi G, Kotre C, Dowling A. Reference
  levels in ptca as a function of procedure complexity. Radiation protection
  dosimetry, 2005;117:54--58.

\bibitem[{Sakano et~al.(2017)Sakano, Iwamoto, Kuribara, Sakamoto, Tajima,
  Hamano, Maruyama, Kikuchi, Tsukamoto, and Kato}]{sakano2017survey}
Sakano T, Iwamoto T, Kuribara T, Sakamoto H, Tajima O, Hamano Y, Maruyama M,
  Kikuchi T, Tsukamoto A, Kato K. Survey of exposure dose during percutaneous
  coronary intervention for chronic total occlusion. Nihon Hoshasen Gijutsu
  Gakkai zasshi, 2017;73:51--56.

\bibitem[{Siiskonen et~al.(2018)Siiskonen, Ciraj-Bjelac, Dabin, Diklic,
  Domienik-Andrzejewska, Farah, Fernandez, Gallagher, Hourdakis, Jurkovic,
  et~al.}]{siiskonen2018establishing}
Siiskonen T, Ciraj-Bjelac O, Dabin J, Diklic A, Domienik-Andrzejewska J, Farah
  J, Fernandez JM, Gallagher A, Hourdakis CJ, Jurkovic S, et~al. Establishing
  the european diagnostic reference levels for interventional cardiology.
  Physica Medica, 2018;54:42--48.

\bibitem[{{STUK}(2016)}]{stukdrl}
{STUK}. P\"a\"at\"os 15/3020/2016: Potilaan s\"ateilyaltistuksen vertailutasot
  kardiologisessa radiologiassa, 2016.

\bibitem[{Suomi et~al.(2019)Suomi, Komar, Sainio, Joronen, Perheentupa, and
  Sequeiros}]{suomi2019comprehensive}
Suomi V, Komar G, Sainio T, Joronen K, Perheentupa A, Sequeiros RB.
  Comprehensive feature selection for classifying the treatment outcome of
  high-intensity ultrasound therapy in uterine fibroids. Scientific reports,
  2019;9:10907.

\bibitem[{{Task Force on Assessment of Diagnostic and Therapeutic
  Cardiovascular Procedures}(1988)}]{aha1}
{Task Force on Assessment of Diagnostic and Therapeutic Cardiovascular
  Procedures}. Guidelines for percutaneous transluminal coronary angioplasty.
  Circulation, 1988;78:486--502.

\bibitem[{Urbanowicz et~al.(2018)Urbanowicz, Olson, Schmitt, Meeker, and
  Moore}]{urbanowicz2018benchmarking}
Urbanowicz RJ, Olson RS, Schmitt P, Meeker M, Moore JH. Benchmarking
  relief-based feature selection methods for bioinformatics data mining.
  Journal of biomedical informatics, 2018;85:168--188.

\end{thebibliography}

%% References copied and pasted from the .bbl file.  Copy and paste over the following two lines.  When using a bibTeX database to create a .bbl file, comment out the following two lines.

\end{document}